\documentclass{article}
\usepackage{file}

\usepackage{times}
\usepackage{soul}
\usepackage{url}
\usepackage[hidelinks]{hyperref}
\usepackage[utf8]{inputenc}
\usepackage[small]{caption}
\usepackage{graphicx}
\usepackage{amsmath}
\usepackage{booktabs}
\title{Mobile Gamer Lifetime Value Prediction via Objective Decomposition and Reconstruction}

\author{
Tianwei Li\footnote{Author to whom any correspondence should be addressed.}
\and
Yu Zhao
\and
Yunze Li\\
\And
Sheng Li\\
\affiliations
TapTap \\
Shanghai, China \\ 
\emails
\{litianwei, zhaoyu, liyunze, lisheng2\}@xd.com
}

\begin{document}

\maketitle

\begin{abstract}
For Internet platforms operating real-time bidding (RTB) advertising service, a comprehensive understanding of user lifetime value (LTV) plays a pivotal role in optimizing advertisement allocation efficiency and maximizing the return on investment (ROI) for advertisement sponsors, thereby facilitating growth of commercialization revenue for the platform.
However, the inherent complexity of user LTV distributions induces significant challenges in accurate LTV prediction.
Existing state-of-the-art works, which primarily focus on directly learning the LTV distributions through well-designed loss functions, achieve limited success due to their vulnerability to outliers.
In this paper, we proposed a novel LTV prediction method to address distribution challenges through an objective decomposition and reconstruction framework.
Briefly speaking, based on the in-app purchase characteristics of mobile gamers, our model was designed to first predict the number of transactions at specific prices and then calculate the total payment amount from these intermediate predictions. 
Our proposed model was evaluated through experiments on real-world industrial dataset, and deployed on the TapTap RTB advertising system for online A/B testing along with the state-of-the-art ZILN model.

\end{abstract}

\section{Introduction}

Real-time bidding (RTB) advertising has emerged as a predominant monetization strategy for digital platforms with substantial user bases, including social media networks, e-commerce platforms, and cyber communities. 
To maximize commercial revenue, such platforms are essential to optimize advertising efficiency and enhance the return on investment (ROI) of advertisers. 
One of the key components is understanding the value of the users, including the intention to purchase and the capability of payment.
The user lifetime value (LTV), which quantifies the total economic value of a customer relationship to the merchant, serves as one of the fundamental marketing metrics for user value evaluation.
Practical applications of the LTV metric not only enable effective allocation of sponsored items to improve platform advertising performance \cite{vanderveld2016engagement}, but facilitate high-value user acquisition (UA) strategies \cite{su2023data} that significantly benefit the ROI of advertisers.
Consequently, accurate prediction of user LTV has become an indispensable capability for advanced RTB advertising systems.

\begin{figure}[t]
	\centering
	\includegraphics[width=\linewidth]{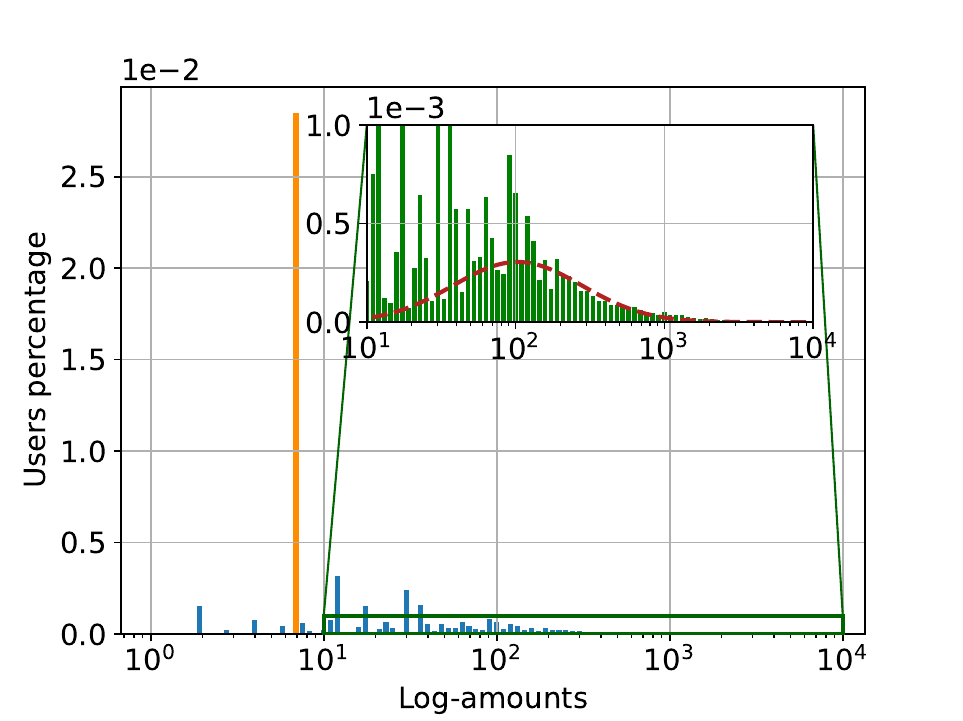}
	\caption{
    The distribution of the cumulative payment amount of mobile gamers.
    In lower amount range, a substantial portion of paying users only purchase the initial bundle, as shown by the orange bar.
    For amount above 10 (depicted by the green bars in the zoomed figure), the distribution roughly follows a log-normal pattern (represented by the red dashed curve) with numerous outliers. 
    }
	\label{fig_2}
\end{figure}

For third-party community platforms with RTB service, however, it is challenging to predict user LTV, primarily due to data-related issues.
One of the difficulties lies in the quality and completeness of user transaction data, including purchase frequency and payment amount. 
These data points are usually either incomplete or inaccurate for third-party platforms, as the acquisition of the data points is highly dependent on the data callback from the advertisers.
Such challenge has been partially addressed by the optimized cost-per-action (oCPA) advertising technique \cite{zhu2017optimized}.
Specifically, advertisers are required to provide accurate in-app purchase data of the users to the advertising platform to optimize the bidding strategies, balancing user acquisition volume and advertising ROI. 
Nevertheless, the inherent nature of user spending patterns presents another significant hurdle. 
The distribution of the cumulative payment amount of users, i.e., the distribution of user LTV, exhibits extreme skewness, with a small fraction of high-spend users contributing prominently to the total payment amount \cite{fader2005rfm}, resulting in substantial challenges for LTV prediction.
Conventional regression models with mean squared error (MSE) loss \cite{toro1968test} are particularly sensitive to high-spend outliers, hardly capturing the skewed distribution. 
Recent works \cite{wang2019deep} and \cite{zhang2023out} proposed deep probabilistic models utilizing zero-inflated log-normal (ZILN) loss to quantify LTV with certain uncertainty across both low and high spenders. 
However, the assumption that user LTV follows a log-normal distribution may not hold true in the mobile gaming industry (as illustrated in Fig. \ref{fig_2}), potentially limiting the improvement of prediction accuracy.

In this paper, we proposed CALTV (\textbf{C}umulative \textbf{A}mount LTV), a deep neural network (DNN) architecture integrating decomposition and reconstruction networks (DRNets), to handle the distribution challenges in predicting user in-app payment amount in the mobile gaming industry. 
Our methodology generally involved a two-stage process: objective decomposition, followed by reconstruction from predicted sub-objectives.
Specifically, the total in-app payment amount of each user was first decomposed into individual transactions with specific prices. 
A DNN model was then trained to predict sub-objectives representing the number of transactions for each price category. 
Finally, the total payment amount of each user was computed by aggregating these sub-objective predictions.
A key innovation of our CALTV model was the design of loss function, which focused on optimizing the prediction of transaction counts for each price category rather than directly fitting the cumulative payment amount. 
This design effectively mitigated the impact of outliers in the LTV distribution which typically hinder the performance of conventional regression methods \cite{wang2019deep,zhang2023out}.
As a result, our approach achieved notable improvements in both offline experiments on our internal production dataset and online A/B testing on our RTB advertising system.

The remainder of the paper is structured as follows. 
Section \ref{sec_related_work} reviewed previous research works on LTV modeling. 
The architecture and methodology of our proposed CALTV model was elaborated in Section \ref{sec_proposed_method}.
Section \ref{sec_experiment} presented the implementation and results of the data experiment.
Section \ref{sec_conclusion} concluded the contributions and key findings of this work.

\section{Related Works}
\label{sec_related_work}
User lifetime value modeling has been studied in academia and industry for decades, from statistical models to machine learning models and neural network models.
Methods \cite{schmittlein1987counting,fader2005counting} based on “Buy ’Til You Die” (BTYD) framework were proposed first to differentiate commercial value of users by modeling the transaction process (Buy) and the dropout process (Die) during the entire lifetime.
The early model \cite{schmittlein1987counting} was developed based on the number and timing of historical transactions by fitting a Pareto distribution for purchase probability and a negative-binomial distribution (NBD) for purchase frequency.
The beta-geometric/NBD (BG/NBD) model \cite{fader2005counting} optimized the computational consumption by replacing the Pareto distribution with a BG distribution for user purchase probability.
However, the BTYD family models have limitations in LTV prediction as the models focus on predicting consumption habits instead of the spending amount of the users. 
\cite{fader2005rfm} expanded the BYTD method with the Recency-Frequency-Monetary (RFM) value to directly estimate the LTV of users. 
This stochastic model applies a Pareto distribution for the most recent purchase (recency) and a NBD distribution for the transaction counts (frequency), which adopts the BYTD framework, followed by a gamma-gamma distribution for the spending amount per transaction (monetary).

As data scale increased with the development of digital platforms in recent years, large numbers of handcrafted embedding features began to be employed in LTV prediction, resulting in superior performance of the machine learning-based and deep learning-based models.
\cite{vanderveld2016engagement} and \cite{chamberlain2017customer} divided users into several cohorts based on historical purchase data and introduced a two-stage random forest model to first recognize paying users by a binary classifier and then predict the spending amount for each predicted payer. 
\cite{wang2019deep} optimized the complexity of the two-stage tree-based models by introducing the deep learning framework into the LTV modeling problem.
As the distribution of the user LTV was illustrated as a zero-inflated log-normal format, a corresponding mixture loss was proposed to simultaneously learn the purchase propensity and monetary value of the users.
Although the application is limited by the restrictive distribution assumption, this deep learning-based model brings an innovation methodology to LTV modeling, resulting in more attention to the investigation of distribution patterns.
The subsequent ODMN method \cite{li2022billion} simplified the LTV distribution modeling problem by dividing the imbalanced distribution into several balanced sub-distributions, each fitted by a distribution expert.
And the game whale detector \cite{zhang2023out} proposed a multitask framework with two LTV experts integrating the ZILN loss to predict the monetary values of low and high spenders, respectively.
Nevertheless, the LTV modeling still needs to be customized due to the variation of the LTV distribution in different business contexts.

\begin{figure}[t]
	\centering
	\includegraphics[width=\linewidth]{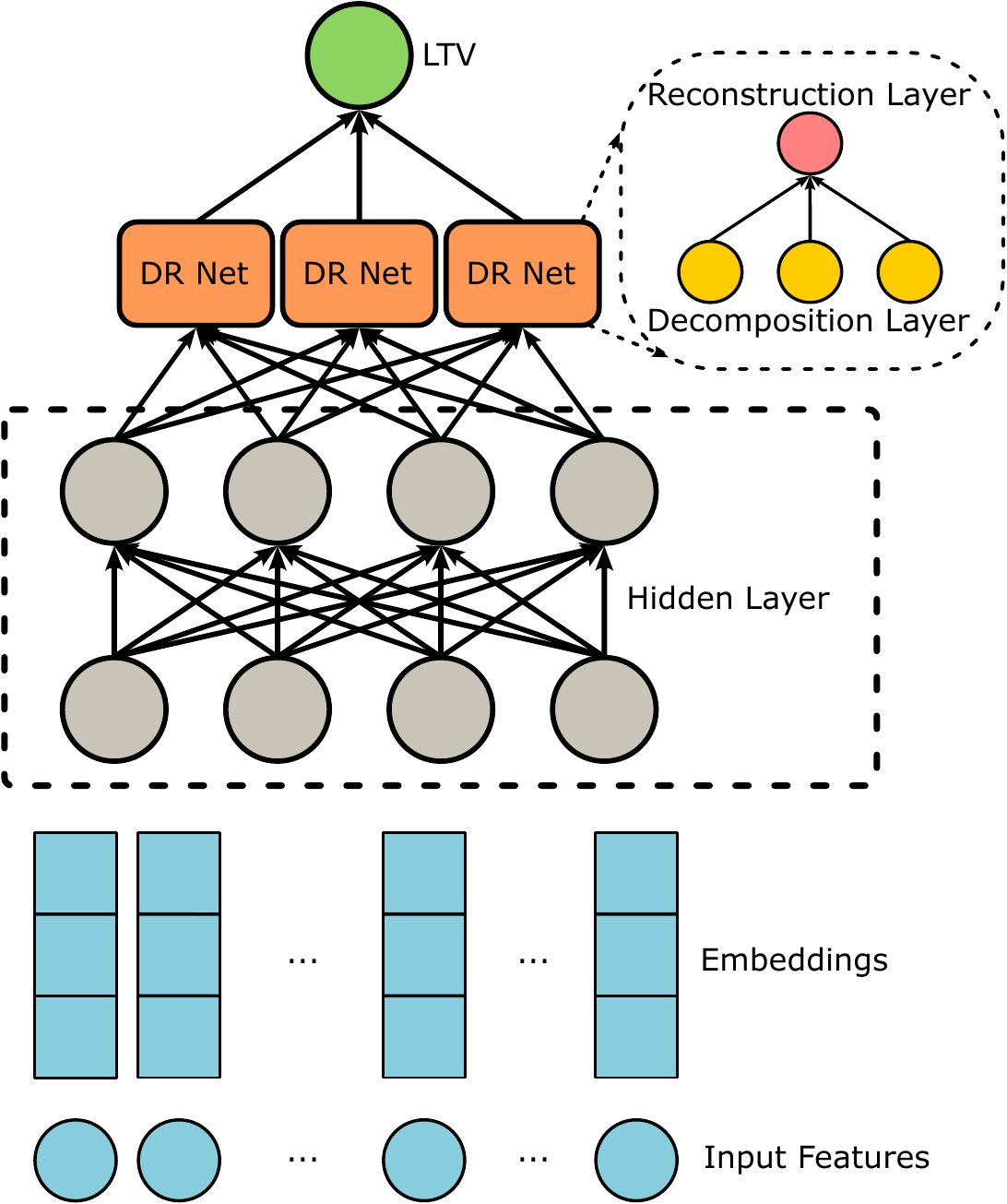}
	\caption{The schematic representation of the CALTV model. A DNN structure with an embedding layer, several fully connected hidden layers, a layer of decomposition and reconstruction nets to predict the sub-objectives (transaction counts), and an additional reconstruction layer as the output layer to generate the final objective (LTV) by aggregating the intermediate sub-objectives.
        }
	\label{fig_caltv}

\end{figure}

\section{Methodology}
\label{sec_proposed_method}
In this section, the proposed LTV prediction model, CALTV, was presented in detail. 
The model, as shown in Fig. \ref{fig_caltv}, was built on a basic DNN structure with two key components: a layer of decomposition and reconstruction nets (DR nets) to predict the number of transactions for each price category and an additional reconstruction layer to aggregate these intermediate predictions to generate the final LTV.

\subsection{LTV Decomposition}
\begin{figure}[t]
	\centering
	\includegraphics[width=\linewidth]{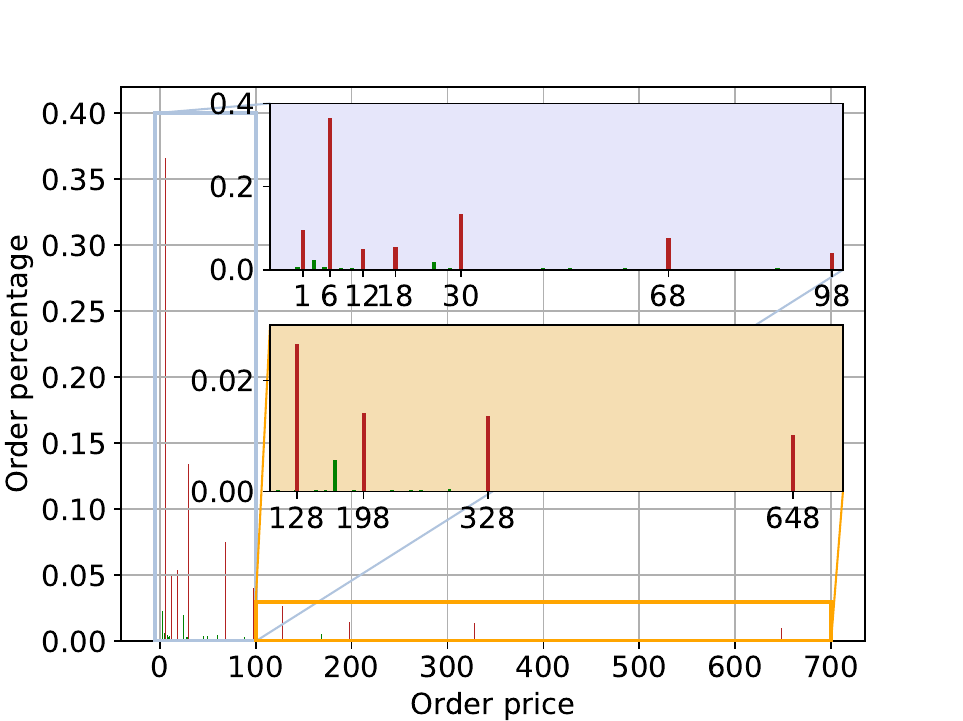}
	\caption{
    The distribution of model game in-app transaction orders price.
    Significant portion of transactions occur at specific prices such as CN¥1, CN¥6, CN¥12, CN¥18, CN¥30, CN¥68, CN¥98 represented by the red bars in the top zoomed figure, and CN¥128, CN¥198, CN¥328, CN¥648 in the bottom figure, while transaction orders with other price constitute a negligible portion.
    }
	\label{fig_3}
\end{figure}

The LTV model deployed in the RTB advertising system of our mobile gaming community was designed to predict the total payment amount $V_{u,i}$ of a user $u$ to a sponsored mobile game $i$ within a specified time period $T$.
This total payment amount objective, which was applied as a unitary number in existing models, is generally composed of the amount of each transaction order, i.e.
\begin{equation}
\label{eq1}
    V_{u,i} = \sum_{n=1}^{N} V_{n,u,i}
\end{equation}
where $V_{n,u,i}$ denotes the spending amount of the $n$-th transaction order and $N$ is the total number of transactions for the user-item pair $(u,i)$.
Next, these $N$ transaction orders could be categorized into $M$ groups according to the respective amount, i.e.,
\begin{equation}
\label{eq2}
    V_{u,i} = \sum_{m=1}^{M} C_{m,u,i} V_{m,u,i}
\end{equation}
where $C_{m,u,i}$ denotes the number of transaction orders with amount $V_{m,u,i}$, and $\sum_{m=1}^{M} C_{m,u,i} = N$.
This representation in Eq. (\ref{eq2}) is particularly appropriate in the mobile gaming in-app payment scenario, as mobile gamers frequently purchase promotion bundles that offer better value than random recharges.
These bundles are typically standardized with common prices such as CN¥6, CN¥30, CN¥68, and CN¥128 (equivalent to \$1, \$5, \$10, and \$20), as shown in Fig. \ref{fig_3}.
As a result, by Eq. (\ref{eq2}), the LTV prediction task was decomposed into a limited $M$ number of subproblems, each focusing on predicting the transaction order counts $C_{m,u,i}$ for a specific price category $V_{m,u,i}$.

\subsection{Transaction Counts Prediction}
\label{sec_oc_prediction}
\begin{figure}[t]
	\centering
	\includegraphics[width=\linewidth]{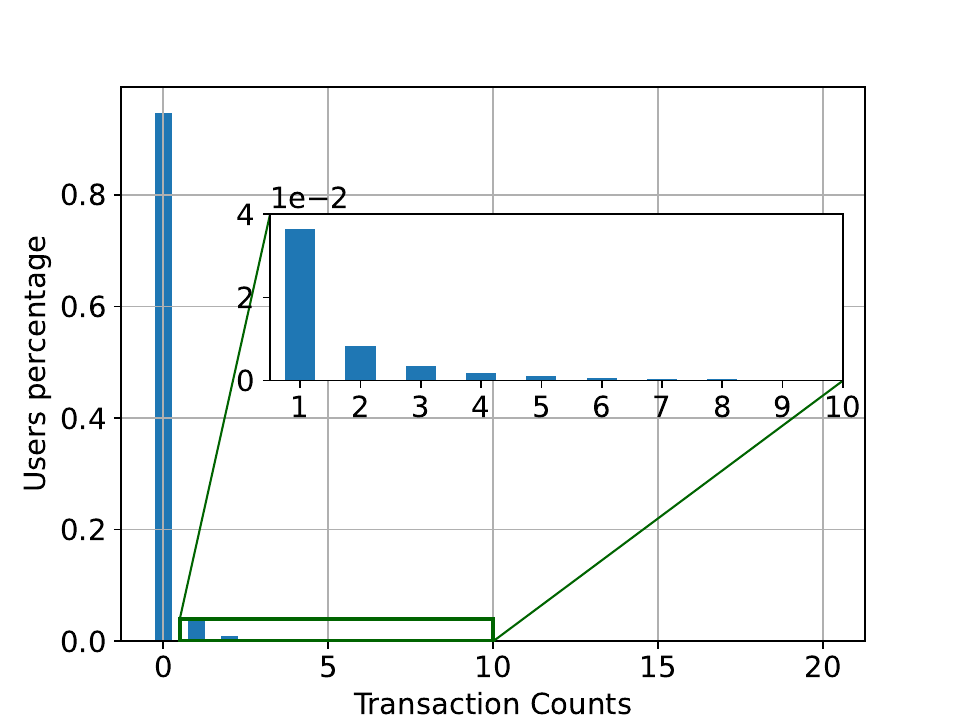}
	\caption{
    The distribution of transaction counts of the mobile gamers.
    Nearly 95\% of the mobile gamers are non-spenders, while more than 95\% of the paying users make fewer than 5 transactions.
    }
	\label{fig_5}
\end{figure}

Predicting transaction counts is much more tractable than predicting LTV directly, as the distribution of transaction counts (primarily ranging from 0 to 5 as shown in Fig. \ref{fig_5}) is more intensive than the LTV distribution.
However, it is still challenging to predict transaction counts directly by regression model with mean square error (MSE) loss due to the sparsity of payment behaviors.
\cite{fader2005counting} addressed the impact of label sparsity and distribution outliers by quantifying the uncertainty of transaction counts with negative binomial regression (NBR).
This approach, however, underperformed in the mobile gaming context as the number of high-amount transactions deviated from the NB distribution, partially due to business promotion of the high-price bundles.
To address the limitations, we applied a simple yet effective logistic regression model based on a DNN structure to predict the number of transaction orders for each price category.
Specifically, for a given price category $m$, the training label $y_{m,u,i}$ of each sample $(u,i)$ in the dataset $\mathcal{S}_D$ was first generated by truncating the transaction counts at an upper limit to remove the outliers and relatively homogenize the distribution, i.e., 
\begin{equation}
\label{eq3}
    y_{m,u,i}=\min(C_{m,u,i},\widehat{C}_m) 
\end{equation}
where $\widehat{C}_m$ denotes the upper limit of the transaction counts for category $m$.
Subsequently, a multi-nominal logistic regression model was employed to predict the probability $p_{c,m,u,i}$ of user $i$ purchasing $c$ numbers of game $i$ bundle with price $V_{m,u,i}$, for $c \in [1,\widehat{C}m]$, by optimizing the cross-entropy loss function:
\begin{equation}
\label{eq_loss_count}
    \mathcal{L}_{m} = -\sum_{(u,i) \in \mathcal{S}_D} \sum_{c=1}^{\widehat{C}_m} y_{c,m,u,i}\log(p_{c,m,u,i})
\end{equation}
where $y_{c,m,u,i}=1$ if $y_{m,u,i} = c$ else $0$.
Finally, the predicted transaction counts for the price category $m$ was represented by the expected value $\hat{y}_{m,u,i}$ of the logistic regression model, i.e.,
\begin{equation}
\label{eq5}
    \hat{y}_{m,u,i} = \sum_{c=1}^{\widehat{C}_m}p_{c,m,u,i}c
\end{equation}
This method effectively mitigated the impact of outliers and addressed the distribution of transaction counts for each price category in a consistent process.

\subsection{Objective Reconstruction}
As the transaction counts for each price category were predicted using the logistic regression model introduced in SubSec. \ref{sec_oc_prediction}, the final LTV objective could be reconstructed through Eq. (\ref{eq2}).
Specifically, the model simultaneously predicted transaction counts for all $M$ numbers of price categories by minimizing the aggregated cross-entropy losses:
\begin{equation}
    \label{eq_loss_ltv}
    \begin{split}
    \mathcal{L}_{LTV} &= \sum_{m=1}^{M} \mathcal{L}_{m} \\
    &= - \sum_{(u,i) \in \mathcal{S}_D} \sum_{m=1}^{M}\sum_{c=1}^{\widehat{C}_m} y_{c,m,u,i}\log(p_{c,m,u,i})
    \end{split}
\end{equation}
where $\mathcal{L}_{m}$ was computed by Eq. (\ref{eq_loss_count}).
We consolidated these $M$ numbers of sub-objective predictions with a single DNN-based model sharing the input embeddings and the hidden layers, as illustrated in Fig. \ref{fig_caltv}, to simplify the model structure and optimize computational efficiency.
The LTV prediction $\hat{V}_{u,i}$ was finally represented by the total expected payment amount across all price categories, which was obtained by substituting the number of transaction orders $C_{m,u,i}$ in Eq. (\ref{eq2}) with the predicted one $\hat{y}_{m,u,i}$ in Eq. (\ref{eq5}):
\begin{equation}
    \hat{V}_{u,i} = \sum_{m=1}^{M} \left( V_{m,u,i}  \sum_{c=1}^{\hat{C}_m}p_{c,m,u,i}c \right). 
\end{equation}

\section{Experiments} 
\label{sec_experiment}
In this section, we evaluated and analyzed the performance of the proposed CALTV model through data experiment on an internal production dataset collected from our RTB advertising system, benchmarking against the state-of-the-art ZILN model \cite{wang2019deep} and the game whale detector ExpLTV \cite{zhang2023out}.
Our model was also deployed on the TapTap RTB advertising system for online A/B testing along with the ZILN baseline.

\subsection{Dataset and Implementation}
The experiment dataset contained nearly 30 million conversion samples (i.e., clicking the ``acquire" button of the sponsored items) collected over a 12-month consecutive period from TapTap advertising system.
Each sample consisted of numerical and categorical features processed from user profile, and sequential features from user historical behaviors, in addition to the sponsored item features and the request context features.
The labels of each sample were derived from the in-app purchase data provided by the advertisers. 
Specifically, each payment data point was first attributed to the corresponding conversion event, and then the amounts paid within time window $T$ after the conversion occurred were aggregated as the $T$ period LTV label for the corresponding sample.
In this experiment, we set $T$ to a 24-hour period to optimize the short-term value of the users.
In addition, we also calculated the labels representing the transaction counts of each significant price in Fig. \ref{fig_3} for the implementation of the proposed CALTV.
Specifically, given the significant prices in ascending sequence $0<V_1<V_2<\cdots<V_M \in \mathcal{S}_V$, an order at price $v$ was classified into the transaction category with the significant price $V_m \in \mathcal{S}_V$ if $V_m \leq v <V_{m+1}$.

The proposed model and the baselines were initially trained on the first 10 months of the samples.
To simulate the online application scenarios, each model was then incrementally fine-tuned and evaluated on the following 2 months of the samples day by day.
Specifically, for each day $D$, we fine-tuned the model checkpoint on the previous day $D-1$ using the samples of day $D$, and then applied this day $D$ checkpoint to predict the LTV of the samples on day $D+1$.
Numerical metrics, such as the area under curve (AUC) and the prediction bias, were applied to evaluate the discrimination and precision of each model.

\subsection{Results and Discussion}
\begin{figure}[t]
	\centering
	\includegraphics[width=\linewidth]{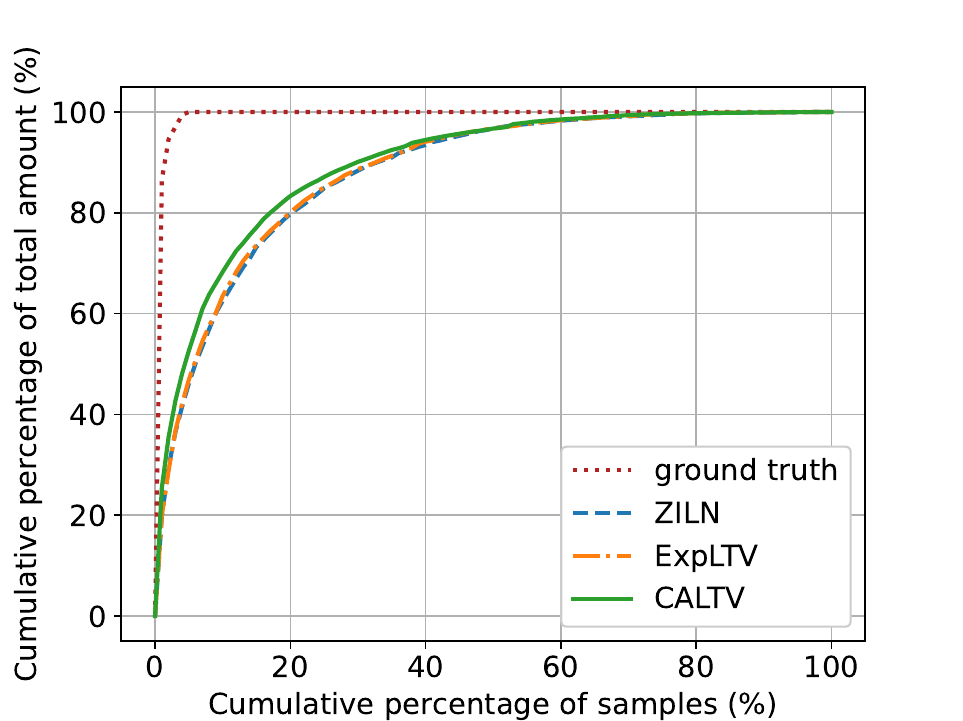}
	\caption{
     Comparison of the Lorenz curves between baseline models and the proposed CALTV model. 
     Each curve was plotted by first ranking the evaluation samples in descending order according to the LTV predictions of each model, and then accumulating the payment amount of the samples from the highest ranking one to lowest ranking one.
     The ground truth curve was derived in the same approach but ranking based on actual payment amount rather than model predictions.
    }
	\label{fig_6}
\end{figure}

First, we evaluated the user value discrimination capability for each model using the AUC metric.
However, due to the continuous nature of the LTV labels, the conventional approach to AUC calculation, which measures the probability that a classifier ranks the positive samples higher than the negative ones \cite{huang2005using}, was not applicable to LTV models.
Inspired by the computation of the Gini coefficient in \cite{wang2019deep}, we redefined the AUC for LTV models as the area under the Lorenz curve (AULC), which represents the average of the cumulative percentage of total payment amount against the cumulative percentage of total samples ordered by model predictions in descending sequence,
\begin{equation}
\label{eq_8}
    \mbox{AULC} = \frac{1}{K}\sum_{k=1}^{K} \frac{V_k^{\mathrm{cum}}}{V^{\mathrm{tot}}} 
\end{equation}
where $V_k^{\mathrm{cum}} = \sum_{i=1}^{k} V_i$ and $V^{\mathrm{tot}} = V_K^{\mathrm{cum}}$ denote the cumulative payment amount until the top $k$-th sample group and the total payment amount of all the evaluation samples, respectively.
This metric is similar to the AUC metric of binary classifiers, with higher values indicating better discrimination of the model.
The Lorenz curves and the corresponding AULC by Eq. (\ref{eq_8}) of our proposed model and the baselines are presented in Fig. \ref{fig_6} and Table \ref{tab_A}, respectively.
Apparently, the Lorenz curve of our proposed model (depicted in solid green in Fig. \ref{fig_6}) consistently outperforms the curves of both baselines (the dashed blue and the dash-dotted orange curves) in any range of the evaluation samples.
Specifically, our proposed model improved the AULC metric by more than 1.4\% in absolute value compared to both baselines, demonstrating superior discrimination of high-value users.

\begin{figure*}[t]
	\centering
	\includegraphics[width=\textwidth]{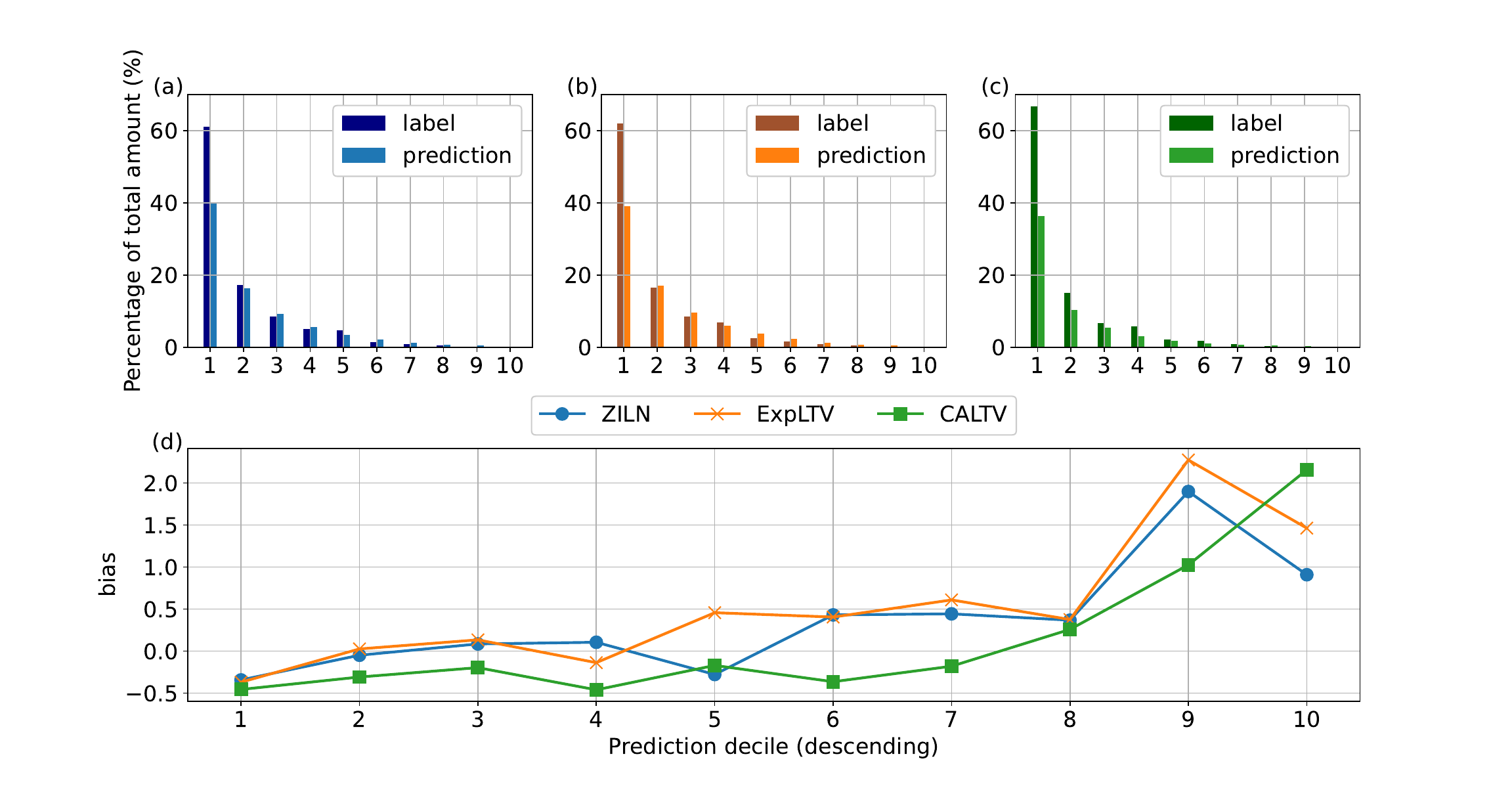}
	\caption{
    Comparison of the decile biases between baselines and our proposed model.
    The predicted payment amount of (a) the ZILN model, (b) the ExpLTV model, and (c) the CATLV model are presented respectively versus the actual payment amount for each prediction decile in descending sequence.
    The corresponding bias across these 10 groups are presented in (d).
    }
	\label{fig_7}
\end{figure*}

Next, we evaluated the variance in prediction bias, as this metric had a significant impact on the performance of the oCPA-based \cite{zhu2017optimized} real-time bidding strategy in our advertising system.
Specifically, prediction biases with small variance can be easily fixed with a well-designed oCPA bidding risk factor, while big variance may induce failure in high-value user acquisition.
To assess the bias variance of each model, we first evenly divided the samples into $K$ groups according to the corresponding predictions of each model in descending sequence, and then calculated the variance of the group biases (GBiasVar):
\begin{equation}
\label{eq_gbias_var}
    \mbox{GBiasVar} = \frac{1}{K}\sum_{k=1}^K \left( \mbox{Bias}_k - \widetilde{\mbox{Bias}} \right)^2
\end{equation}
where $\mbox{Bias}_k=\frac{\hat{V}_k - V_k}{V_k}$ denotes the predicted bias of group $k$ with the predicted value $\hat{V}_k$ and the label $V_k$, and $\widetilde{\mbox{Bias}}=\mbox{med}_{k=1}^{K}(\mbox{Bias}_k)$ denotes the median of the biases, respectively.
Figures \ref{fig_7}(a)-(c) present the predicted payment amount of each baseline and our proposed model in 10 groups along with the actual payment value for each group, respectively.
Apparently, all models underestimated the top first-tier samples and overestimated the lowest two deciles, which may be due to the challenge of modeling the extremely skewed distribution of the payment amount.
Fig. \ref{fig_7}(d) presents the corresponding decile biases of each model.
It is obvious that the curve of the decile biases of our proposed model was smoother than those of both baselines, particularly for the top 80\% samples.
Group bias variances calculated using Eq. (\ref{eq_gbias_var}) are presented in Table \ref{tab_A}.
Although the overall GBiasVar of our proposed model was higher than the baselines due to the overestimation in the lowest two deciles, the variance for the top 80\% samples was significantly optimized by more than 40\%, which was advantageous for the oCPA-based bidding strategy.

\begin{table}
	\caption{Model performance on AULC and GBias variance}
	\label{tab_A}
	\begin{tabular}{ c c c c }
		\toprule
		Model & AULC & GBiasVar & GBiasVar (top 80\%) \\
		\midrule
		CALTV & \textbf{0.902} & 0.7353 & \textbf{0.0469} \\		
		ZILN & 0.886 & \textbf{0.4033} & 0.0824 \\		
		EXPLTV & 0.888 & 0.5802 & 0.1017 \\
		\bottomrule
	\end{tabular}
\end{table}

Finally, we deployed our proposed CALTV model on the TapTap RTB advertising system for online A/B testing along with the ZILN baseline.
Briefly speaking, we split the online traffic equally into two halves, and for each advertising request, given the short-term ROI target (24-hour period in our online application), the user value predicted by the corresponding model was applied directly in real-time bidding using the oCPA strategy to optimize the marketing ROI of each sponsored item.
As a result, the CALTV model achieved over 20\% promotion in the 24-hour period ROI compared to the ZILN baseline, demonstrating its superiority in practical application in mobile game marketing.

\section{Conclusion} 
\label{sec_conclusion}
In this paper, we proposed the CALTV model to predict the LTV of mobile gamers.
Our model addressed LTV prediction challenges, including skewed distributions and extreme outliers, through an innovative objective decomposition and reconstruction framework.
Specifically, a two-stage approach was employed on a DNN structure to first predict the transaction counts at specific prices and then aggregate these intermediate predictions to generate the LTV objective.
The effectiveness of our model was demonstrated through offline data experiments on our internal production dataset and online A/B testing on our RTB advertising system.

\bibliographystyle{unsrt}
\bibliography{reference}

\end{document}